# What Output-Equivalence Oracles Miss: An Empirical Study of Equivalence-Invisible Bug Fixes in Quantum Transpilers

Furqan Nasir[a],[b], Muhammad Arif Shah[a], Iftikhar Alam[a]
[a] City University of Science and Information Technology (CUSIT), Peshawar, Pakistan
[b] National University of Computer and Emerging Sciences (FAST-NUCES), Islamabad, Pakistan.
Corresponding author: Furqan Nasir (furqannr@gmail.com)

## Abstract

Quantum compilers face the test oracle problem, judged by an output-equivalence oracle: the compiled circuit must compute the same unitary as the original, modulo global phase and a qubit-layout permutation. This oracle, by construction, checks only that semantic map, not the circuit's own layout, permutation, or phase records: a defect there, or in a fixed-seed run's determinism, can pass unseen though the record is public. This empirical software engineering study of quantum transpiler correctness uses repository mining to measure how often this happens in real merged compiler fixes: a systematically identified corpus of Qiskit transpiler bug-fixes, classified by an independently dual-coded, source-validated fault-manifestation taxonomy. Nineteen of 68 fixes (28%, 95% Wilson CI 19–40%) repair faults invisible to this equivalence screen, even one augmented with compilation-validity, circuit-quality, and performance checks, and an extended 104-fix corpus over a wider window holds at the same rate with a tighter interval (29/104, 27.9%, CI 20–37%). A conservative floor remains even restricted to the one unconditionally equivalence-invisible channel (a corrupted layout or permutation record): 10 of 68 fixes (15%, CI 8–25%) beneath the 28% headline. The gap is not Qiskit-specific: it replicates in tket (7/21, 33%), with Cirq smaller but consistent. We detected no systematic differences on five inexpensive PR-level characteristics (19 vs 49, underpowered alone; the extended 29-vs-75 corpus tightens every interval toward zero). This class dominates the invisible set, concentrating at representation-boundary crossings. We release the corpus, codebook, and coding artifacts for quantum software testing research. Here we only measure it.

## 1. Introduction

A compiler for quantum circuits, or transpiler, rewrites an abstract circuit into one a specific device can run. It maps gates onto a native set and a limited qubit connectivity, through a sequence of layout, routing, translation, optimization, and scheduling passes. Deciding whether such a rewrite is correct is an instance of the test-oracle problem [1]: given an input and an output, by what criterion do we judge the output right? In practice, a transpiler is judged, by the compiler's own regression-test suite and by the formal-verification tools built for it, against output equivalence: the compiled circuit must compute the same unitary as the original, up to a global phase and a permutation of the qubit layout. Qiskit's own transpiler-pass tests assert exactly this, typically via an operator-equivalence check between a pass's input and output circuit (e.g. test/python/transpiler/test_optimize_1q_decomposition.py). The two formal-verification efforts built specifically for this compiler, Giallar and CertiQ, verify passes against this same output-preservation property [2], [3]. Efficient equivalence-checking techniques make that criterion practical to apply at scale.

Output equivalence is a natural and necessary criterion, but it is not a complete one. By construction, an output-equivalence oracle inspects only the map from inputs to outputs. A defect can leave that map intact while still corrupting something the compiler is obliged to get right. That something might be a layout or permutation property that downstream code relies on, the determinism of a fixed-seed run, or a global phase that becomes observable once the circuit is placed under a controlled operation. Each of these is a public attribute of the compiled circuit, not a hidden one; the output-equivalence oracle simply does not compare it. Such a fault escapes output equivalence and the compilation-validity, circuit-quality, and performance checks usually bundled with it, because none of those signals changes either. The fix that repairs it is, in our terms, equivalence-invisible relative to that reference screen, not undetectable by any conceivable check.

Whether this blind spot matters is an empirical question, not a conceptual one. If equivalence-invisible faults are rare curiosities, the incumbent oracle is a reasonable approximation and the gap is academic. But if they are common, a substantial share of real compiler defects is being validated against a criterion that cannot see them, and the field's confidence in transpiler correctness rests on a screen with a known hole of unknown size. Empirical work on quantum software has already established that compiler and transpiler components are a leading source of defects, and that many such defects manifest as unexpected outputs rather than crashes [4], [5]. Existing quantum-testing techniques (from differential and metamorphic testing to fuzzing and equivalence-modulo-inputs [6], [7]) all reason, one way or another, over program outputs. None of them, though, measures how much of the real defect stream escapes a semantic-equivalence-centric correctness criterion of the kind this paper defines as its reference screen (Table 1). That measurement is this paper's subject.

We measure it directly. We systematically mine merged Qiskit transpiler bug-fixes over a bounded recent window, through a multi-source discovery protocol, screen each candidate against an explicit scope rule, and classify the surviving fixes by the channel through which their fault could be detected. Two coders apply a frozen codebook independently and adjudicate disagreements against it. A subsample of labels is checked against each fix's own source diff and regression test, so the classification is anchored to source-level evidence, not to how a fix happens to be described. We then ask whether the resulting rate reproduces in other, independently developed compilers, whether equivalence-invisible fixes can be told apart from ordinary ones by cheap surface signals, and what kinds of fault recur in the invisible channels.

The measurement is stark. Roughly a quarter to a third of merged Qiskit transpiler bug-fixes repair faults an output-equivalence oracle cannot observe: 19 of 68 fixes in the primary corpus (28%, 95% Wilson confidence interval 19–40%), holding at 29 of 104 (27.9%, CI 20–37%) on an extended corpus drawn from a wider window. We report both as headline measurements of the rate. The gap replicates in tket, an independently developed compiler (7 of 21, 33%), with Cirq's smaller sample pointing the same direction. We detected no systematic differences between equivalence-invisible and observable fixes on the five inexpensive PR-level signals examined, with small effect sizes throughout, on both the primary corpus (19 vs 49) and an extended corpus drawn from a wider window (29 vs 75); the latter tightens every interval toward zero. Its largest component, corrupted contract or permutation metadata, concentrates where a value crosses an internal representation boundary — a pattern we confirm by reading each fix's own diff.

These findings answer four research questions:

- **RQ1.1.** How often do merged Qiskit transpiler bug-fixes fall into channels a reference output-oracle screen cannot detect?
- **RQ1.2.** Does the equivalence-invisible rate replicate in an independently developed compiler?
- **RQ1.3.** Can equivalence-invisible fixes be distinguished from observable ones by cheap surface signals such as size, churn, or merge latency?
- **RQ1.4 (exploratory).** What fault mechanisms recur among the equivalence-invisible fixes?

In answering them, this paper contributes:

- a study-specific fault-manifestation taxonomy and an operational definition of the observability gap, stated relative to an explicit reference oracle screen;
- the first direct measurement of the observed corpus fraction falling in that gap in a real compiler's merged-fix stream, established on two corpora that agree closely: a systematically assembled, exhaustively screened, independently coded and source-validated frame (n = 68) and an extended, wider-window corpus (n = 104) that reproduces the rate at a tighter interval;

- a replication in a second, independently developed compiler (tket), with a third (Cirq) pointing the same way at smaller scale, establishing that the gap is not an artifact of a single codebase;
- a negative result, confirmed and tightened on a second, larger corpus, that equivalence-invisible fixes show no detectable difference from observable ones on the cheap surface signals we measured, weighing against the obvious low-cost workaround;
- a diff-verified characterization of the recurring fault mechanisms in the invisible channels; and
- a released corpus, frozen codebook, and complete coding and adjudication artifacts.

Our claim is deliberately bounded. We establish that the gap exists, measure its size in this corpus and window, and show it is not explained by the surface features we tested. We do not claim that any particular oracle closes it. Whether a purpose-built family of oracles can observe these faults is out of scope here (§3.4): this paper is a measurement, and the measurement stands independently of any proposed remedy. The corpus is also deliberately small and exhaustively hand-validated, rather than large and automatically classified. Every fix is screened against an explicit scope rule, independently dual-coded, and, for a subsample, checked against its own source diff. This trades the cross-repository breadth of large-N mining studies for per-fix ground truth. That trade is the design this question requires, because the question is what a specific oracle class structurally misses on real fixes, not how many repositories exhibit it (§8).

# 2. Background

## 2.1 The Qiskit transpiler

Qiskit compiles an abstract circuit to a target backend through a staged pass manager [8]. The standard stages are layout, routing, translation, optimization, and scheduling, bundled into optimization levels 0–3. Two properties matter here. The passes that actually run depend on the circuit, backend, configuration, and the exact Qiskit revision. And transpilation may permute qubits via inserted routing operations [3], so a transpiled circuit equals the original only modulo a layout permutation.

## 2.2 Fault types and their manifestation

We distinguish what kind of defect a fix addresses from how, and indeed whether, it shows up in an observable signal, because the gap between these two is this study's central finding. The correctness group contains functional faults (a compilation failure) and semantic faults (a different unitary or output). The quality group covers circuit-quality faults, such as a worse two-qubit count or a larger depth. The performance group covers compilation-performance faults.

Crucially, two defects of the same fault type can manifest in very different observable signals, and some manifest only in signals the output-equivalence oracle never inspects. We therefore use a fault-manifestation taxonomy: the channel through which a defect could be detected (Table 1). The reference correctness screen is an augmented oracle suite: the output-equivalence oracle together with compilation-validity, circuit-quality, and performance checks. We reserve the term output-equivalence oracle for the semantic-output comparison itself.

**Table 1. Study-specific fault-manifestation taxonomy (detection channel). The reference screen is an augmented oracle suite: the output-equivalence oracle plus compilation-validity, circuit-quality, and performance checks.**

| Manifestation channel | Observable signal | Augmented suite sees it? |
|---|---|---|
| output_semantic | different unitary/output state | Yes (output-equivalence oracle, within width) |
| compilation_failure | crash, exception, invalid circuit | Yes (compilation-validity check) |
| circuit_quality | worse two-qubit count or depth | Yes (quality check) |

| Manifestation channel | Observable signal | Augmented suite sees it? |
|---|---|---|
| performance | slower / more memory-hungry compilation | Yes, if over the performance screen |
| transpiler_contract_or_metadata | wrong layout/permutation property (TranspileLayout, final_layout, routing_permutation) with applied output still correct | No, invisible to the whole suite, unconditionally |
| global_phase | tracked global phase becomes wrong or is dropped, with the compiled unitary otherwise correct | No, unless the phase is later made observable (e.g. under a controlled operation) |
| determinism_or_reproducibility | output varies across fixed-seed runs, often only under a specific (e.g. noisy) target | Only with a determinism oracle on that target |

### 2.3 The oracle problem for transpilation, briefly

Deciding whether a transpiled circuit is correct is an instance of the quantum oracle problem. A correctness check compares what a transpiler exposes as output against an expected result. The field's standard notion of that comparison, the one this paper adopts as its reference screen, is semantic output equivalence modulo global phase and qubit-layout permutation (§1). That definition discards phase and layout information by construction, not because the information is hidden. The layout and permutation contract, the tracked global phase, and fixed-seed reproducibility are all ordinary, public attributes of the object a transpile call returns. A fault that corrupts one of them, while leaving the compiled unitary correct, therefore passes the reference screen unseen. This is not because no check could ever compare that attribute, but because the specific, widely used equivalence notion this paper measures against does not. This is the boundary the manifestation taxonomy (Table 1) makes explicit, and it is the boundary this paper measures the size of. Whether and how such faults can be detected, by an oracle built specifically to compare the missing signal, is a separate, detection-side question that this paper does not take up.

## 3. Related Work

### 3.1 Empirical studies of compiler and quantum-software bugs

Empirical characterizations of defects in quantum software and compilers are still comparatively rare. Bugs4Q curates 36 real, manually validated Qiskit bugs across Terra, Aer, Ignis, and Aqua, with buggy and fixed versions and reproduction tests. It is positioned as a reusable historical-bug benchmark, not a prevalence study of any particular fault property [9]. Broader empirical work characterizes bug and quality-attribute patterns across quantum software at scale [4]. A separate large-scale mining study links technical debt to fault occurrence across 118 open-source quantum repositories [10]. Recent methodological guidance for empirical quantum-software-testing studies argues for more rigorous reporting of sampling frames, coding protocols, and reliability statistics in this literature [11] — a gap this paper's methodology (§4) is deliberately built to close. None of these studies asks how often a real, merged fix repairs a fault that a specific, named class of oracle (output equivalence) cannot observe. That is the question this paper's mining study answers.

### 3.2 Mining-study and coding-reliability methodology in software engineering

Repository-mining studies that classify defects from titles, descriptions, and linked artifacts routinely face a construct-validity question: does a coded label track the underlying defect, or only its written description? The methodological literature on inter-coder agreement in empirical software engineering treats this as a first-class concern. It argues that reliability coefficients (Cohen's kappa, Krippendorff's alpha) must be computed on raw, pre-adjudication labels, and reported alongside a description of blinding and adjudication, rather than asserted from adjudicated labels alone [12]. We follow this guidance directly (§4.4-4.6): both coders are blinded to each other and to prior agreement statistics, raw labels are retained even after adjudication, and kappa is reported with confidence intervals rather than as a bare point estimate. For interpreting the resulting values we use the conventional benchmark scale [13]. Our primary channel-taxonomy kappas, across the three Qiskit dual-coding rounds (§4.6, §5.2), range 0.62-0.96 and fall in the substantial-to-almost-perfect band. The cross-SDK replication kappas (§5.3) are reported and interpreted separately against the same scale: tket's is comparable (0.77, substantial), while Cirq's smaller, exploratory sample yields a lower, moderate figure (0.52), which we do not fold into the range above. Our methodology departs from a fully qualitative coding study at the source-validation step (§4.7). Because the corpus is code, not interview or survey data, a subsample of coded labels can be checked directly against source-level evidence: the fix's own diff and regression test — an option most qualitative-coding guidance does not have.

### 3.3 Quantum software testing

A structured literature search across major scholarly databases (IEEE Xplore, SpringerLink, and arXiv, with the full review additionally covering the ACM Digital Library, ScienceDirect, Scopus, Web of Science, and Google Scholar) identified seven primary studies directly concerned with quantum compiler and quantum-software testing. All seven target bug detection or provide a bug benchmark, and none targets prevalence of faults invisible to a specific oracle class. QDiff performs differential testing across quantum software stacks (Qiskit, Cirq, PyQuil), using statistical comparison of output distributions [6]. MorphQ applies metamorphic testing with generated Qiskit programs and quantum-specific transformations, exposing real bugs through metamorphic-relation violations [7]. QuteFuzz structurally fuzzes quantum compilers with randomly generated circuits containing control flow and subcircuits, detecting crashes and compiler/simulator inconsistencies [14]. QEMI adapts equivalence modulo inputs to quantum programs: it removes quantum-control-flow dead code to generate variants, then compares output distributions across Qiskit, Q#, and Cirq [15]. Giallar and CertiQ take a formal-verification approach instead. Giallar verifies 44 of 56 Qiskit passes across 13 versions using symbolic execution and SMT solving [2]. CertiQ mostly-automatically verifies 26 of 30 Qiskit compiler passes against contracts using quantum circuit calculus, explicitly checking that verified passes still pass Qiskit's own regression tests [3]. Across all seven studies, the oracle strategy is either statistical output comparison, metamorphic relations, crash detection, equivalence-modulo-inputs, or formal semantic preservation, and every one of them reasons about the compiled output or an equivalence relation over outputs. None of these oracle strategies is designed to observe contract or layout metadata, tracked global phase, or fixed-seed determinism as a distinct signal. That is precisely the boundary this paper's mining study measures.

A supplementary citation check (2026-09-06) surfaces two further, closely related tools from the same research group, not among the seven primary studies above but worth noting for completeness. LintQ statically flags likely bugs in Qiskit programs, finding that most of what it detects is missed by prior static tools [16]. QITE cross-checks Qiskit against three other platforms using a crash oracle plus an equivalence oracle over exported assembly [17]. Both still reason about program state or output, rather than the contract and phase signals this paper targets. Also relevant, though aimed at a different question, is a recently proposed test oracle that validates a quantum program's own probabilistic output states more reliably under repeated measurement [18]. That line of work improves the correctness of output-level oracles themselves. This paper instead asks what is not detected by the semantic-equivalence-centric reference screen defined in Table 1, regardless of how reliably any given output-level oracle is implemented. A related but distinct line of work designs the equivalence-checking algorithms themselves: efficient black-box procedures for deciding whether two quantum programs compute equivalent, identical, or unitarily related maps [19]. Better equivalence-checking algorithms make the output-equivalence oracle in Table 1 cheaper or more scalable to apply, but they do not change what that class of oracle can see by construction. That is the boundary this paper measures, rather than tightens.

The gap this paper measures also has a direct analogue in classical compiler testing, where the field faced a version of the same oracle problem decades earlier. Translation validation, for instance, checks that each individual compilation preserves the source program's semantics, rather than proving the compiler correct once and for all [20] — the same output-preservation stance the output-equivalence oracle in Table 1 takes toward transpilation. A related classical technique, equivalence modulo inputs, generates program variants provably equivalent to the original with respect to a specific test input, then compiles both and compares outputs on that input. It has found real miscompilations in production C compilers this way [21], and is the classical counterpart to QEMI's quantum adaptation [15]. GCC and LLVM's own miscompilation bugs have been characterized the same way: an empirical study of real bug reports shows how they distribute across compiler phases and optimization levels [22], a classical parallel to this paper's own mechanism-level breakdown of where equivalence-invisible faults concentrate (§5.5). Nondeterminism is a distinct, well-studied failure mode in classical test suites too. An empirical study of flaky tests across large software projects found that failures recur from a set of root causes dominated by concurrency, timing, and test-order dependencies, rather than by defects in the code under test [23]. This is the same class of defect this paper's determinism/reproducibility channel targets in the transpiler. None of these four studies addresses quantum compilation. Together, though, they show that an output-only correctness criterion missing non-output properties, and nondeterminism escaping a fixed test oracle, are not quantum-specific problems: they recur wherever a black-box, input-output notion of correctness is the field's default. Quantum transpilers therefore provide a contemporary setting in which to study a broader software-engineering question: how often semantic-preservation checks fail to cover auxiliary correctness contracts maintained by transformation systems.

Two tools published as this study was being finalized reinforce the same point from the most recent literature. QSPE follows the differential-testing principle to enumerate skeletal quantum programs and validates them by statevector comparison rather than measurement sampling, reporting 708 miscompilations across quantum libraries, 81 of them acknowledged by the Qiskit team [24]. QuCheck brings property-based testing to Qiskit, with the developer-stated properties still adjudicated over the executed program's output [25]. Both are recent, effective additions to the output-based family. Neither inspects the transpiler's own contract, phase, or determinism metadata as a distinct signal, which remains the boundary this paper measures.

### 3.4 Positioning

This paper's contribution is orthogonal to all seven of the primary studies above. Equivalence checking, differential and metamorphic testing, fuzzing, equivalence-modulo-inputs, and formal pass verification each aim at finding bugs or certifying input-output behaviour more effectively. Each, by construction, reasons about the compiled output or an output-level equivalence relation. We instead ask how often a real, merged transpiler fix repairs a fault that lies entirely outside that observation boundary, and whether that rate is particular to one compiler or one project's testing culture. No primary study in the preliminary search quantifies this, consistent with the search's own finding that dedicated regression-testing or test-selection frameworks for quantum compilers are largely absent from the literature. A companion paper (Nasir et al., in preparation for Software Testing, Verification and Reliability) takes the fault classes measured here and asks the complementary question: whether a fault-class-matched oracle family can detect them on real historical fixes. That is a detection contribution, and is out of scope here.

## 4. Methodology

This section specifies the mining and coding protocol in full, since the validity of every result in §5 rests on it. We report the sampling frame and search strategy, the inclusion and exclusion rules with their exact counts, how the codebook was developed and frozen, and who coded the corpus and under what blinding. We also report how disagreements were adjudicated, and how agreement was computed and should be interpreted. Finally, we report how a subsample of labels was checked directly against source, and how surface characteristics were extracted for RQ1.3.

### 4.1 Sampling frame and search strategy

The observation window is merged pull requests to the Qiskit repository between 2025-05-08 and 2026-07-01, spanning the 2.x transpiler line. Candidates were pooled from two rounds: a 24-fix adjudicated seed, and a 44-fix expansion. The expansion was drawn through three widening discovery passes over four source streams: the mod: transpiler changelog label, a title-and-

keyword sweep over pass names and channel terms, release-note seeds, and an official release-and-PR audit. Label-based discovery alone is incomplete. A whole-library scan of one year of merged fixes found that the mod: transpiler and Changelog: Fixed labels surface only 28 of 39 distinct transpiler bug-fixes we could independently identify. These labels under-represent the global-phase channel the most: labels alone caught 3 of the 9 global-phase fixes in that scan. The three discovery passes were therefore: (i) the changelog label itself, (ii) the keyword sweep, which recovers fixes tagged mod: circuit or left unlabelled, and (iii) a semantic review of titles and descriptions, which recovers cases with no matching label or keyword at all. We report the resulting corpus as a lower bound on the true population of in-scope fixes, not a census. All identified candidates were exhaustively screened, but candidate discovery cannot itself be guaranteed exhaustive. Accordingly, 19 of 68 is the observed fraction in this systematically assembled corpus, not an unbiased estimate of the complete population of Qiskit transpiler fixes.

## 4.2 Inclusion/exclusion criteria and screening flow

A candidate is in scope only if it is a merged pull request that fixes a real defect in the transpiler or a directly adjacent high-level-synthesis component. Feature additions, performance-only enhancements with no documented defect, documentation-only changes, and pure refactors are excluded, as are backport duplicates of an already-counted primary fix. Figure 1 traces the resulting funnel. An existing human-coded baseline of 26 pull requests was combined with 59 newly identified candidates, screened against the rules above. Of these, 44 were included and 15 excluded: backport duplicate, 4; already present in the baseline or an explicitly superseded duplicate, 5; feature or enhancement rather than a defect fix, 2; performance enhancement rather than a correctness fix, 2; documentation-only, 1; refactor with no documented product defect, 1. The 44-candidate pool was assembled incrementally: 39 candidates carried forward from an earlier screening pass, plus 5 further distinct in-scope defects identified through continued mining (PRs #16249, #16246, #16151, #16154, #15137), reaching 44. This nominal 26 + 44 = 70-candidate corpus then went through a second, independent eligibility re-audit of all 44 candidates, checking PR-number duplication against the baseline, known backport duplication, and component scope. All 39 carried-forward candidates and all 5 newly added candidates were confirmed in scope, with no replacements required. Separately, a scope gate applied to the original 26-item baseline removed two items found on closer reading not to be in-scope defects (one non-bug feature/behaviour change, one outside the transpiler component), leaving 24. The final analytic corpus coded and reported throughout this paper is therefore 68: the 24-item adjudicated baseline plus the 44 re-audited additions. The full per-candidate screening log, with every decision and its stated reason, is released as data/mining_validation/screening_log_70.csv, and the re-audit log as included_candidate_audit_70.csv.

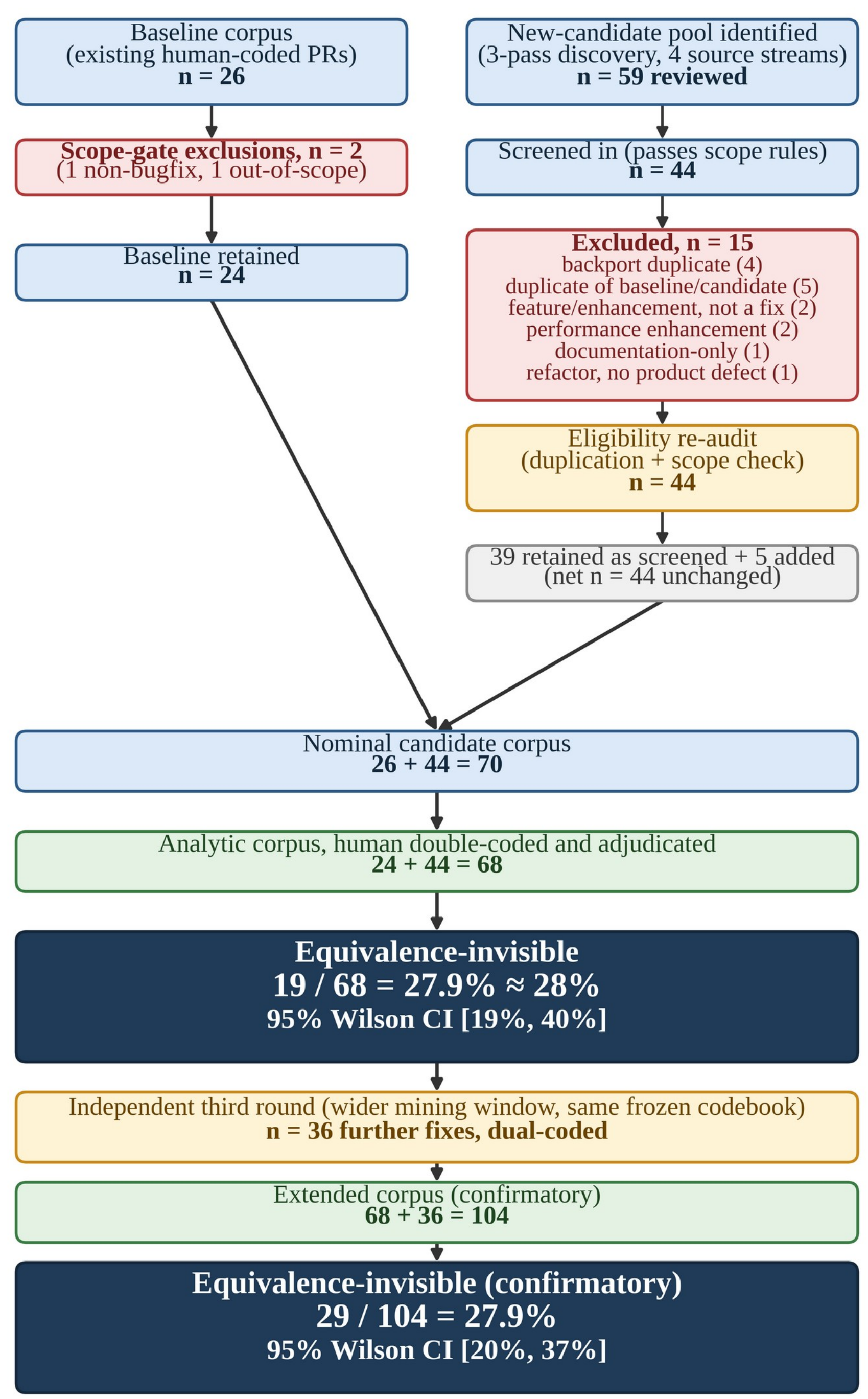

**Figure 1. Screening flow from the nominal 70-candidate corpus to the 68-fix analytic corpus, with exact stage counts and exclusion reasons, extended to show the independent third round (36 further fixes, dual-coded against the same frozen codebook) that forms the confirmatory 104-fix corpus (§5.1). Source data: data/mining_validation/screening_log_70.csv, candidate_queue_70.csv, included_candidate_audit_70.csv, labels_final_68.csv, labels_final_104.csv.**

## 4.3 Codebook development and freezing

Fixes are classified by a seven-channel manifestation taxonomy (Table 1, §2.2): output_semantic, compilation_failure, circuit_quality, performance, contract/metadata, global phase, and determinism/reproducibility, plus an explicit exclude value for out-of-scope candidates. The codebook also states worked adjudication rules, per channel, for the boundary cases coders most often disagree on. A crash or exception is compilation_failure, regardless of any accompanying metadata symptom (Rule 1). A change that alters only compile time or effort, with an identical output circuit, is performance; a change to the output circuit itself is circuit_quality (Rule 2). A change that an output-equivalence oracle, modulo global phase and qubit permutation, would not detect is contract/metadata rather than circuit_quality, even when it looks like an allocation or quality issue on the surface (Rule 3). The codebook (data/mining_validation/CODEBOOK_v2_FROZEN.md) was frozen before the 44-fix expansion round began coding, so no channel definition could be adjusted in response to fixes seen partway through the expansion. This mirrors, for the qualitative coding, the same freeze-before-evaluation discipline that the numerical anti-leakage controls apply to thresholds and seeds elsewhere in this line of work.

### 4.4 Coders, training, and blinding

Two human annotators with quantum-software experience coded the corpus independently: the first author (R1) and a second, independent annotator (R2). Each coder worked from a worksheet containing only the PR identifier, title, URL, and linked-issue reference. Per the coding instructions, a coder reads the pull request's title, description, and linked issue, and reads the code diff only when that text is ambiguous. This way, classification tracks how a triager would actually encounter the fix, rather than requiring a full code review. R2 was blind to R1's labels, to the adjudication history, and to all prior agreement statistics for every round. Blinded worksheets are released as evidence of this separation (e.g. tket_worksheet_BLINDED.csv, cirq_worksheet_BLINDED.csv, and the separately recorded human_worksheet_44_R1.csv/R2.csv for the Qiskit round). For every fix, each coder recorded exactly one manifestation channel and a binary observable-by-output-oracle judgment, which the channel determines, so recording it independently is a built-in consistency check. Each coder also recorded a confidence level (high, medium, or low) and a free-text justification, mandatory whenever confidence was low or the item was excluded. Coder identity and provenance are recorded in data/mining_validation/RATER_PROVENANCE.md. R1 is the first author and so has no separate declaration. Signed declarations from the independent coders are released for R2 (Muhammad Atif Saeed: declarations/Coder_Declaration_Atif.pdf) and for the third volunteer introduced in §4.5 (R3, Muhammad Sajjad Saleem: declarations/Coder_Declaration_Sajjad.pdf).

### 4.5 Independent coding rounds and adjudication

Coding proceeded in two rounds. In the seed round, both coders independently classified the 24-item baseline. In the expansion round, both independently classified the 44 newly screened items, blind to the seed round's adjudicated labels as well as to each other. Disagreements in both rounds were resolved through a structured adjudication step, not by simple majority or by deferring to either coder. Both coders reviewed the disputed item together against the specific codebook rule that governed the boundary (Rules 1-3 above, or the relevant channel definition), then consulted the underlying PR, issue, and diff as needed, and recorded a final label together with the reason on an adjudication sheet (data/mining_validation/adjudication_decisions.csv). As an auxiliary, non-decisive cross-check, two independently prompted large-language-model passes (over the same frozen codebook) were also solicited for each disputed item and shown alongside the coders' raw labels during adjudication. They were never used to decide a disagreement, and never pooled into any reported agreement statistic; §4.6 reports why this separation matters. Raw per-coder labels were never overwritten by the adjudicated value: both are retained in the released data so a reader can recompute agreement independently. A third volunteer independently coded the full Qiskit corpus against the same frozen codebook as an additional check. Those labels are released (rater3_sheet.csv), but by design are not pooled into the headline R1-R2 statistic reported below. The same R2 also independently coded the later 36-fix confirmatory round (§5.1, §5.2), so the headline R1-R2 reliability comparison uses a consistent pair of coders across all three rounds. The third volunteer (R3) did not take part in that round.

### 4.6 Agreement computation and interpretation

Agreement is reported as Cohen's kappa on the raw, pre-adjudication labels, computed separately for the binary observable-by-output-oracle judgment and for the seven-class manifestation channel, with bootstrap 95% confidence intervals (data/mining_validation/compute_kappa_analytic.py for the seed; scripts/score_worksheet.py for the expansion). On the 24-item seed, binary kappa is 0.667 (95% CI 0.338-0.917, 83.3% raw agreement) and channel kappa is 0.619 (95% CI 0.360-0.835, 70.8% raw agreement). On the 44-item expansion, binary kappa is 0.86 (95.5% raw agreement). Following conventional benchmarks for kappa (Landis and Koch, 1977), both are in the substantial-to-almost-perfect range. We report the raw agreement percentage alongside kappa throughout, so a reader is not dependent on any single convention for interpretation. As a transparency cross-check, not as inter-rater reliability, we additionally classified the seed corpus with two independently prompted LLM-assisted passes using the same frozen codebook. Notably, the two LLM passes agreed with each other (binary kappa 0.92) more than the two human coders agreed with each other (0.667). We read this as evidence of shared model priors, rather than of genuine, reproducible codebook application. This is why we base the headline reliability figure solely on the human double-coding, and report the LLM passes only as an auxiliary, clearly labelled comparison (§4.5).

### 4.7 Source validation of labels (construct validity)

Title-and-description-based coding is efficient, but it raises a construct-validity question: does a coded label track the actual fault, or only how it is described? To check this, we validated a subsample of 16 of the 68 coded labels directly against the fix's own source diff and, where one exists, its regression test. These checks operate on source evidence rather than on description. For each label we confirmed, from the fix commit itself, that the fault genuinely occupies the coded manifestation channel. For part of the subsample we went further still, rebuilding both the fixed and parent revisions from source and reproducing the fault directly, or re-running it through a released reproduction script. Eleven fixes were checked this way: ten of the nineteen equivalence-invisible fixes, plus one fix correctly coded observable as a positive control. All eleven matched their coded channel. A second, symmetric audit checked five observable-labelled fixes for the opposite failure mode: a fix wrongly coded observable when it was actually invisible. All five were confirmed genuinely output-visible. Across all sixteen source-checked fixes, no coded label was overturned. Whether these same faults can subsequently be detected by an oracle built for their channel is a separate, detection-side question, taken up in the companion paper (§3.4) rather than here — these checks establish only that the label is correct. The full per-fix breakdown and source citation is in data/mining_validation/label_source_validation.csv and summarised in Table 2 below.

**Table 2. Source-validated labels: coded channel vs. source-confirmed channel (16 of 68 fixes checked, 16/16 match). Each label was confirmed against the fix's own source diff and regression test, with a subset additionally rebuilt from source.**

| PR | Coded channel | Observable? | Source-confirmed channel |
|---|---|---|---|
| #14603 | contract_metadata | no | contract_metadata |
| #14919 | contract_metadata | no | contract_metadata |
| #14956 | global_phase | no | global_phase |
| #16215 | global_phase | no | global_phase |
| #16402 | global_phase | no | global_phase |
| #15024 | contract_metadata | no | contract_metadata |
| #15040 | determinism | no | determinism |
| #14765 | output_semantic | yes (control) | output_semantic |
| #14730 | determinism | no | determinism |
| #16237 | determinism | no | determinism |
| #16201 | global_phase | no | global_phase |

| PR | Coded channel | Observable? | Source-confirmed channel |
|---|---|---|---|
| #13670, #16337, #15626, #15967, #14998 | (observable-labelled, 5-fix false-negative audit) | yes | all confirmed genuinely visible |

### 4.8 Surface-characteristic extraction

For RQ1.3 we asked whether equivalence-invisible fixes could be spotted by cheap surface signals, rather than by a matched oracle. For every one of the 68 fixes we retrieved GitHub metadata: lines added, lines deleted, files changed, commit count, and open-to-merge latency in days (data/mining_validation/pr_characterization_raw.csv). We compared the 19 equivalence-invisible fixes against the 49 observable fixes on each of the five measures, with a two-sided Mann-Whitney U test and Cliff's delta as the effect size. Both were specified a priori, before the comparison was run, at a conventional alpha of 0.05, with no correction search or post-hoc metric selection (data/mining_validation/pr_characterization_summary.csv).

## 5. Results

### 5.1 RQ1.1 — channel distribution of the corpus (n = 68, extended to 104)

Table 3 gives the manifestation-channel distribution of the 68 in-scope, classified merged Qiskit transpiler bug-fixes.

**Table 3. Manifestation channels of the 68 classified Qiskit transpiler bug-fixes. “Reference screen sees it?” denotes detectability by the augmented reference screen defined in Table 1 (output equivalence plus the compilation-validity, circuit-quality, and performance checks bundled with it), not by the output-equivalence oracle alone. The 24-fix seed is double-coded and adjudicated (pairwise Cohen's κ = 0.67 binary). The 44 further fixes are dual-coded (Cohen's κ = 0.86) and adjudicated against the same frozen codebook.**

| Manifestation channel | Count | Reference screen sees it? | Example PRs |
|---|---|---|---|
| compilation_failure | 24 | yes | #13820, #14667, #14998, #15258, #15626 |
| output_semantic | 20 | yes | #13670, #13790, #13874, #16337, #16428 |
| contract / metadata | 10 | no | #13833, #13910, #13945, #14041, #14603 |
| circuit_quality | 4 | yes | #14405, #14869, #15131, #15967 |
| global phase | 5 | no | #14956, #15943, #16201, #16215, #16402 |
| determinism / reproducibility | 4 | no | #14730, #14763, #15040, #16237 |
| performance | 1 | yes (if over screen) | #16476 |

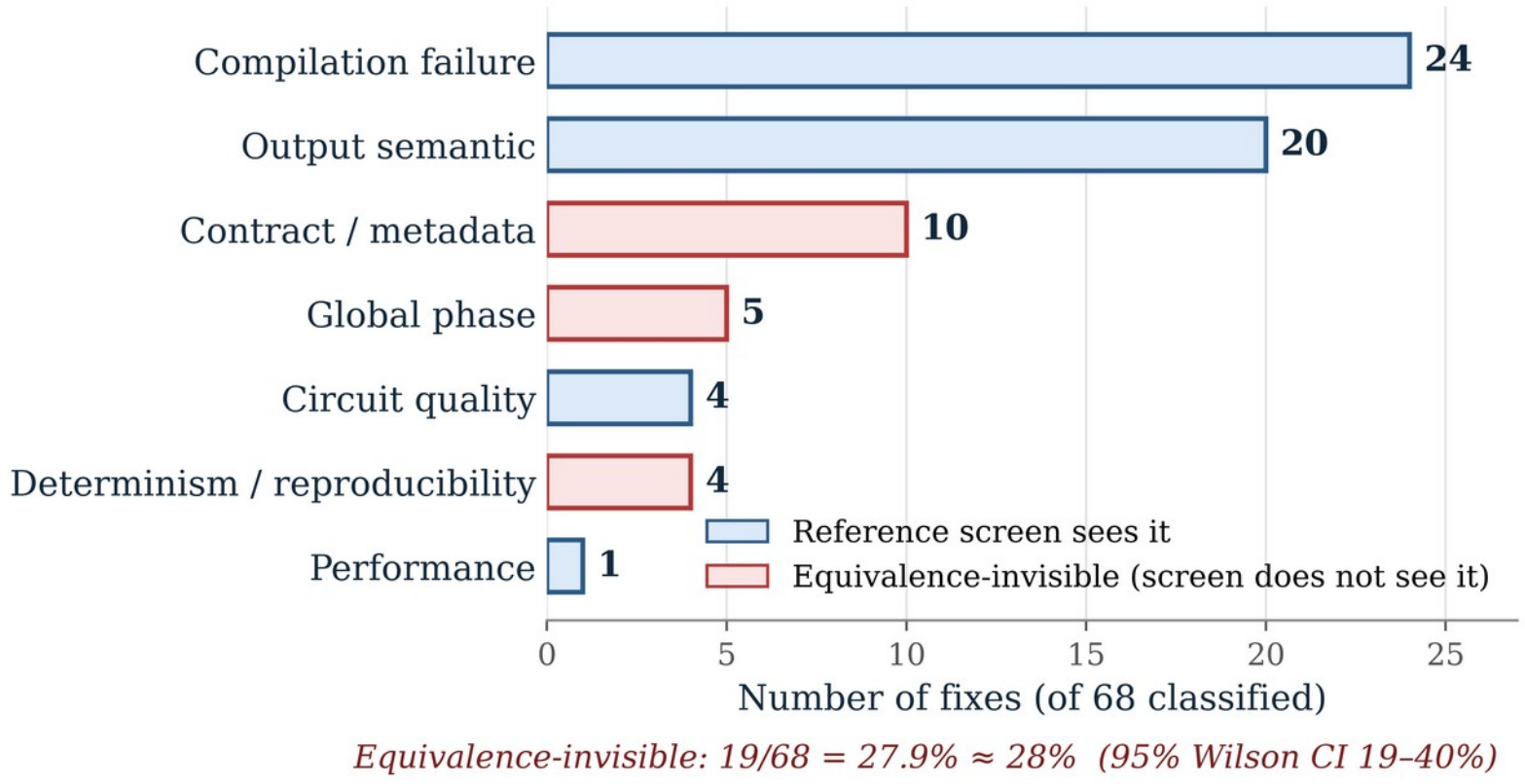

**Figure 2. Manifestation-channel distribution of the 68-fix Qiskit corpus (Table 3), coloured by reference-screen visibility. Source data: data/mining_validation/labels_final_68.csv.**

*19 of 68 classified fixes (≈28%, 95% Wilson CI 19–40%) fall in channels a black-box output-equivalence oracle cannot observe (Figure 2). Of the three invisible channels, only contract/permutation metadata is unconditionally so (Table 1). Global phase and determinism/reproducibility are invisible only under the narrower conditions Table 1 states. Restricting to that unconditional channel alone gives a conservative floor of 10 of 68 fixes (14.7%, 95% Wilson CI 8–25%) beneath the 28% headline. The wider 104-fix corpus (reported next) gives a similar floor (15/104, 14.4%, CI 9–22%). The sample is recent merged fixes, not a complete census, so the percentages are estimates with the reported 95% Wilson CI rather than population parameters. These intervals quantify binomial uncertainty conditional on the observed corpus and do not account for possible channel-dependent candidate-discovery bias.*

The same rate holds on a larger corpus. An independently dual-coded third round of 36 further fixes, drawn from a deliberately wider mining window (older and more recent history) and coded against the same frozen codebook (Cohen's κ = 0.93 binary, 0.96 channel), extends the classification. Combined with the primary rounds, 29 of 104 fixes are equivalence-invisible (27.9%, 95% Wilson CI 20–37%), agreeing with the 68-fix estimate and narrowing the interval on the larger n. We therefore report the observability rate as a headline result on both corpora. The 68-fix corpus is the more uniformly mined, and carries the source-validation and the per-fix analyses that follow. The 104-fix corpus establishes the same rate on a wider window. That extended round carries the channel classification and, as the next section shows, confirms the RQ1.3 surface-signal comparison on a tighter interval. The diff-verified mechanism coding (RQ1.4) below remains scoped to the 68-fix corpus.

## 5.2 Inter-rater agreement and adjudication

On the 24-item seed, binary observable-by-output-oracle kappa is 0.667 (95% CI 0.338-0.917, 83.3% raw agreement) and seven-class channel kappa is 0.619 (95% CI 0.360-0.835, 70.8% raw agreement). Both are in the substantial range. The wide confidence intervals reflect the small round size rather than weak agreement: at n = 24, a handful of disagreements moves kappa considerably. This is exactly why the study also reports the larger expansion round, rather than resting on the seed alone. On the 44-item expansion, binary kappa rises to 0.86 (95.5% raw agreement), almost-perfect, with a materially tighter interval implied by the larger n. The channel-level (seven-class) disagreements in the expansion round numbered ten. Of those, only two touched the binary observable judgment (PRs #14765 and #15074, both edge-of-scope plumbing changes), and both were adjudicated to the observable side. Every disagreement, in both rounds, was resolved through the structured, rule-referenced adjudication protocol of §4.5, never by a simple majority vote or by either coder's unilateral call.

Reading the two rounds together, agreement is substantial to almost-perfect throughout, and improves from the seed to the larger expansion. The small number of adjudicated cases in both rounds concentrates at genuinely ambiguous boundaries the codebook rules were written to resolve (§4.3), rather than scattering evenly across the corpus. A later third round of 36 fixes, mined from a wider window and dual-coded against the same frozen codebook, reached the highest agreement of any round (binary kappa 0.93, seven-class channel kappa 0.96, 97.2% raw agreement). Its single adjudicated disagreement (PR #11351) was resolved from the fix's own diff, rather than its description. Agreement rises across the three rounds even though the frozen codebook's content never changed. Growing coder familiarity with it is a plausible contributor to that rise, though not one this design can isolate from chance.

As a further sensitivity check, we compared the adjudicated R1/R2 labels against the third volunteer's (R3) fully independent labels for all 68 corpus items (§4.5), even though R3's labels are not pooled into the headline reliability figure. Agreement is high (94.1% raw, Cohen's $\kappa = 0.86$). All four disagreements (PRs #14998, #14869, #14765, #16151) point the same direction: R3 classified the fix as equivalence-invisible where the adjudicated R1/R2 label was observable. Taking R3's labels alone would raise the rate to 23 of 68 (33.8%), not lower it. So the third-coder sensitivity analysis provides no evidence that the headline 28% is driven upward by the R1/R2 adjudication — under R3's independent labels, the fraction is higher rather than lower.

### 5.3 RQ1.2 — cross-SDK replication

To test whether the observability gap is specific to Qiskit, we replicated the mining protocol on two independently developed compilers under the same frozen codebook. The replication evidence rests on an independently dual-coded tket sample ($n = 21$, Cohen's $\kappa = 0.77$), which returns the same order of magnitude as Qiskit. Cirq is reported only as a smaller, exploratory consistency check, not as a co-equal third data point. The reason is specific to its codebase — its transformer bug-fix history is close to exhausted at this scope — and we make it explicit below, rather than leave it as an unexplained small sample.

Both replications were mined from each project's own commit history under the same scope gate as Qiskit (data/mining_validation/CODEBOOK_CROSS_SDK.md), not from Qiskit's discovery streams. tket candidates were harvested from CQCL/tket's compilation-pass history (routing, placement, synthesis/rebase, optimization, scheduling, and gating predicates), 2021-10-15 to 2026-02-10: 26 candidates, of which 21 were in scope (5 excluded). Cirq candidates were harvested from quantumlib/Cirq's transformer history, 2022-01-05 to 2026-06-26: 34 candidates, of which only 10 were in scope (24 excluded, overwhelmingly lint, formatting, typo, and coverage-only changes rather than defect fixes). Both corpora were coded by the same R1/R2 pair as the Qiskit rounds, from blinded worksheets (tket_worksheet_R1/R2.csv, cirq_worksheet_R1/R2.csv), against the identical frozen codebook extended by an explicit SDK-specific channel-mapping note, and adjudicated the same way (§4.5). Table 4 and Figure 3 report the resulting per-compiler rates with their 95% Wilson confidence intervals.

**Table 4. Cross-SDK replication of the equivalence-invisible rate. Qiskit and tket are the replication evidence, each independently human-coded and adjudicated against the same frozen codebook (tket $n = 21$, Cohen's $\kappa = 0.77$). Cirq ($n = 10$, $\kappa = 0.52$) is an exploratory consistency check only, not weighted equally with the other two. See below.**

| Compiler | Invisible / in-scope | Rate | 95% Wilson CI | Dominant invisible channel |
|---|---|---|---|---|
| Qiskit | 19 / 68 | 28% | 19–40% | contract / permutation metadata |
| tket | 7 / 21 | 33% | 17–55% | contract / permutation metadata |
| Cirq | 2 / 10 | 20% | 6–51% | contract / permutation metadata |

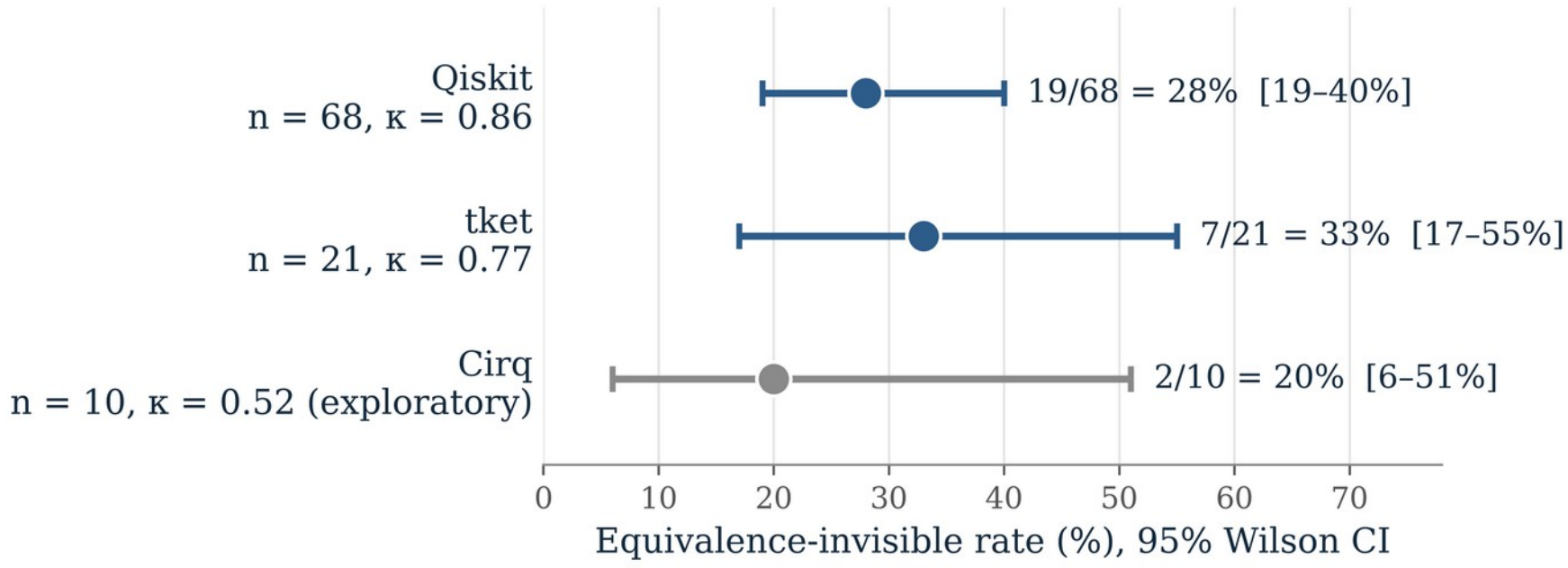

**Figure 3. Cross-SDK replication of the equivalence-invisible rate (Table 4), with 95% Wilson confidence intervals. Qiskit and tket (navy) are replication evidence; Cirq (grey) is an exploratory consistency check.**

Under the predefined scope and mining procedure, Cirq's transformer bug-fix history yielded only 10 eligible fixes, producing a wide interval (6–51%) and moderate inter-rater agreement (κ = 0.52). A fresh check of Cirq's transformer-directory history turned up only one further in-scope candidate beyond the original pool. The items already excluded from that pool are overwhelmingly typo or whitespace housekeeping, so the sample cannot be responsibly grown at this scope. We therefore treat Cirq as exploratory rather than replication evidence. The cross-compiler claim rests on Qiskit and tket, both with substantial-or-better agreement (κ 0.86 and 0.77) and samples large enough to support the reported intervals.

## 5.4 RQ1.3 — surface characteristics (n = 19 vs 49, extended to 29 vs 75)

Table 5 and Figure 4 report the group medians, Mann-Whitney p-values, and Cliff's δ effect sizes with bootstrap 95% confidence intervals, on both the primary and extended corpora.

**Table 5. Equivalence-invisible vs observable fixes: size, complexity, and merge latency (group medians, two-sided Mann–Whitney U) with Cliff's δ and a bootstrap 95% CI (10,000 resamples, seed 42), reported on both the primary 68-fix corpus (n = 19 vs 49) and the extended 104-fix corpus (n = 29 vs 75). No statistically significant difference was detected on either corpus, and every CI brackets zero.**

| Metric | Corpus | Invisible median | Observable median | MW p | Cliff δ | 95% CI (δ) |
|---|---|---|---|---|---|---|
| lines added | 68 | 52 | 54 | 0.66 | +0.07 | [-0.23, +0.37] |
| lines added | 104 | 52 | 45 | 0.90 | +0.02 | [-0.24, +0.27] |
| lines deleted | 68 | 3 | 4 | 0.61 | -0.08 | [-0.39, +0.22] |
| lines deleted | 104 | 3 | 4 | 0.70 | -0.05 | [-0.29, +0.19] |
| files changed | 68 | 3 | 3 | 0.12 | +0.22 | [-0.05, +0.49] |
| files changed | 104 | 3 | 3 | 0.53 | +0.07 | [-0.16, +0.30] |
| commits | 68 | 2 | 2 | 0.53 | -0.10 | [-0.40, +0.22] |
| commits | 104 | 2 | 2 | 0.53 | -0.08 | [-0.32, +0.18] |
| time-to-merge (days) | 68 | 1.18 | 1.15 | 0.74 | -0.05 | [-0.38, +0.27] |
| time-to-merge (days) | 104 | 1.18 | 1.59 | 0.67 | -0.06 | [-0.31, +0.20] |

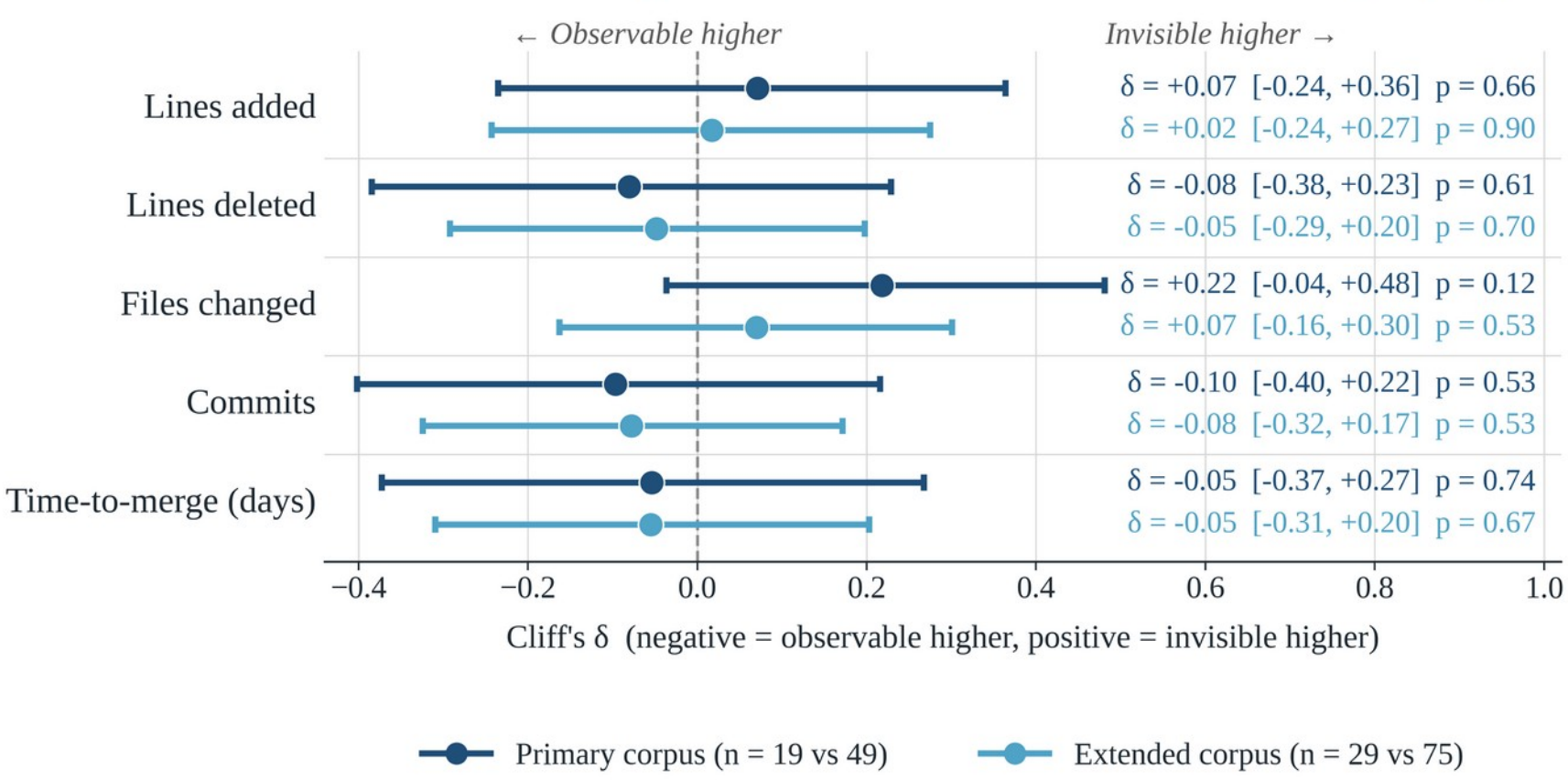

**Figure 4. Effect sizes (Cliff's δ) and bootstrap 95% confidence intervals for the five RQ1.3 surface-characteristic comparisons, on both the primary 68-fix corpus and the extended 104-fix corpus (Table 5). All intervals bracket zero on both corpora, and every interval tightens toward zero on the larger corpus without changing sign.**

On the primary 68-fix corpus, none of the five measures reaches conventional significance (all $p > 0.05$, two-sided Mann-Whitney U), and effect sizes are small throughout (|Cliff's δ| ≤ 0.22). Every bootstrap 95% CI on Cliff's δ brackets zero, including for the largest point estimate (files changed, δ = +0.22, 95% CI [-0.05, +0.49]). So the data are consistent with a true effect of either sign, or with none at all. That comparison is underpowered on its own: at n = 19 vs 49, a true difference of the size hinted at by that largest effect could easily go undetected, and the width of its own CI shows exactly how much room the data leave for that possibility. The extended 104-fix corpus (29 vs 75) addresses this directly, rather than merely gesturing at it. On every one of the five measures the point estimate stays small and near zero, and each interval narrows, most visibly for files changed, whose bound tightens from δ ≤ +0.49 on the primary corpus to δ ≤ +0.30 on the extended one. We therefore read the two corpora together as no detectable difference on the measured signals, not as proof of statistical indistinguishability, and we do not treat this as license to rule out a surface-level triage heuristic entirely. Read as a bound rather than a null, though, the extended-corpus intervals are informative in their own right: across all five signals a true difference larger than a moderate effect is now excluded, including on files changed, the one signal the primary corpus alone could not rule out at that threshold. The reading is thus positive as well as negative, and more decisively so on the larger corpus. These fixes cannot be told apart from ordinary ones by size, churn, or merge latency, which is exactly why separating them calls for a check matched to the fault's own channel, rather than a surface proxy (§6).

## 5.5 RQ1.4 (exploratory) — recurring fault-mechanism categories (dual-coded across four rounds, adjudicated)

Beyond how often a fix is equivalence-invisible, this exploratory analysis asks what kind of fault recurs. An initial open-coding pass classified all 68 fixes into fine-grained mechanism categories from their titles and descriptions. Every assignment was then verified directly against its PR's own source diff, rather than its title or description. For the 19 equivalence-invisible fixes specifically, the R3 volunteer introduced in §4.4-4.5 independently assigned the same eleven categories at diff level, blinded to the original labels: raw agreement was 17 of 19 (89.5%), Cohen's κ = 0.87 (almost-perfect, Landis & Koch 1977). Of the two disagreements, #14603 was adjudicated in favor of the original category. #14939 was later revisited during the full second-coder reconciliation described next, which reclassified it under the codebook's clarified representation-crossing rule. Two contract/metadata fixes are additionally confirmed by verified git archaeology to their introducing commit (#14603 ← cbb4d5d5, "Port ElidePermutations to Rust"; #15024 ← dd8269969, "Port ApplyLayout to Rust").

Beyond that fixed subset, the full 68-fix classification was independently checked by a second coder (R2) across two blind rounds, then reconciled with the first coder (R1) against a revised codebook. In Round 1, R2 coded all 68 fixes from the same eleven categories, blind to R1's labels: raw agreement was 63.2% (43/68), Cohen's κ = 0.590 (κ = 0.620 on the 19 equivalence-invisible fixes, κ = 0.549 on the 49 observable fixes). The 25 disagreements concentrated in two categories, mostly R1's missing_validation_or_contract_check being re-coded as either the catch-all logic_error_pass_specific or api_plumbing_or_serialization. This motivated a revised codebook (v2) stating general elimination rules for exactly these boundaries, rather than fixing individual labels. R2 then recoded all 68 fixes a second time, again blind to the Round-1 labels, using v2: raw agreement was 61.8% (42/68), κ = 0.572, essentially unchanged overall, though the two subsets moved in opposite directions (invisible-subset κ rose to 0.741; observable-subset κ fell to 0.481). Diagnosing this result located a specific ordering flaw in v2: a general representation-crossing rule was tested before two narrower, subsystem-named categories (commutation-checker and global-phase defects), letting the general rule out-compete them on fixes that only incidentally touched those subsystems. A further revision (v3) corrected this by testing the more specific categories first. R1 and R2 then independently reapplied v3 to the 26 fixes disputed across the two rounds, each blind to the other's answer and to any proposed resolution: 23 of 26 (88.5%) converged on the same label, of which 18 corrected R1's original coding. The remaining 3 were resolved by joint discussion of the underlying diff. Table 6 reports the resulting, fully reconciled classification (data/rq14_spotcheck/rq14_mechanism_coding_FINAL_ADJUDICATED.csv); the codebook (data/rq14_spotcheck/RQ1.4_codebook_v3_general_rules.md) is frozen at v3.

**Table 6. Recurring fault-mechanism categories across the 68-fix corpus, dual-coded across two blind rounds and reconciled against a frozen codebook (§5.5). See data/rq14_spotcheck/rq14_mechanism_coding_FINAL_ADJUDICATED.csv.**

| Mechanism category | n | % of 68 | Dominant channel(s) |
|---|---|---|---|
| missing validation / unhandled edge case | 11 | 16.2% | compilation_failure (6) |
| cross-representation / conversion loss | 10 | 14.7% | contract_metadata (4), output_semantic (4) |
| API plumbing / serialization | 9 | 13.2% | compilation_failure (4) |
| logic error, pass-specific (catch-all) | 9 | 13.2% | output_semantic (5) |
| commutation-checker defect | 7 | 10.3% | output_semantic (6) |
| non-deterministic ordering | 5 | 7.4% | determinism (4) |
| global-phase bookkeeping gap | 5 | 7.4% | global_phase (5) |
| unchecked numeric edge case (Rust panics) | 5 | 7.4% | compilation_failure (5) |
| control-flow handling gap | 3 | 4.4% | compilation_failure (2) |
| Rust-port metadata loss (git-confirmed) | 2 | 2.9% | contract_metadata (2) |
| stateful reuse / global cache | 2 | 2.9% | mixed |

The clearest pattern, once every item is checked against its own code and reconciled across coders, is a representation-boundary signature within the contract/metadata channel. Cross-representation/conversion loss and Rust-port metadata loss together account for 6 of the corpus's 10 contract/metadata fixes (60%). Each is a fault surviving the crossing from one internal representation to another (a Python object into a Rust struct, one pass's layout convention into another's), while the compiled output stays correct. This figure survived the full second-coder reconciliation (§5.5) unchanged, but not because nothing in the contract/metadata channel moved. Two of its ten labels changed during reconciliation, one out of this pattern and one into it (#13910 to api_plumbing_or_serialization, #14939 to cross_representation_conversion), leaving the count at 6 of 10 by coincidence of an exact offset, rather than by the pattern being untouched. We report 60% as a coherent, coder-reconciled pattern, rather than a precision claim; the underlying reconciliation, including both of these changes, is documented in data/rq14_spotcheck/rq14_mechanism_coding_FINAL_ADJUDICATED.csv. Two other categories are larger still but orthogonal to this paper's headline: missing validation of edge cases (16.2%) and unchecked numeric edge cases in Rust-side code (7.4%), both already output-visible and reported here for completeness. The eleven categories are not peer abstraction levels. Some name what check is missing (missing validation), others name which component is defective (commutation-checker defect), and one names how the fault was introduced (Rust-port metadata loss). We keep them as reported because the same rule was applied uniformly across the corpus, but the scheme remains a working classification, rather than a validated causal ontology. The reconciliation above covers category assignment under a fixed eleven-category scheme, not whether those eleven categories are the right cut of the space. A further, wholly independent reliability check by new coders, and any coarser regrouping of the categories, are noted as open future work (§8).

## 6. Discussion and Implications

The central finding is quantitative and specific. In this corpus, roughly a quarter to a third of merged Qiskit transpiler bug-fixes (19/68, 95% Wilson CI 19-40%) repair faults that a black-box output-equivalence oracle cannot observe, with a floor of 10/68 (14.7%, CI 8-25%) on the unconditionally invisible contract/metadata channel alone. The same order of magnitude recurs in tket under the identical codebook, with Cirq smaller in scale but pointing the same way. Equivalence-invisible fixes also show no detectable difference from ordinary ones on any of five cheap surface signals we measured, although that comparison (§5.4) is underpowered rather than definitive. None of this depends on why the faults arise. It is a statement about what a specific, widely used correctness criterion misses, established independently of any proposed fix.

**What kind of gap this is.** The mechanism coding (§5.5) adds texture to the headline rate without changing it. It rests on two independent reliability checks: a diff-level check of the 19 equivalence-invisible fixes by a third volunteer (κ = 0.87), and a full-corpus, two-round second-coder reconciliation against a codebook revised twice in response to what the disagreements showed (§5.5). Contract/metadata faults, the largest equivalence-invisible class, appear to include a representation-boundary component, consistent with prior verification work's observation that quantum-data conversions between internal representations are error-prone and warrant explicit correctness guarantees [3]. Six of ten arise where a fault survives the crossing from one internal representation to another (a Python object into a Rust struct, one pass's layout convention composed into another's), while the compiled output stays correct — the largest identifiable pattern in that channel. This is not a claim that Rust rewrites are unusually buggy. The corpus corroborates only two fixes back to a verified introducing commit, both of which happen to be Rust ports, so the pattern should be read as a coherent, moderately-evidenced hypothesis, rather than a settled cause. What it does suggest is a mechanism: a representation crossing is exactly the situation in which output correctness and metadata correctness can be tested by two entirely different code paths. A change that preserves one therefore has no structural reason to preserve the other.

**Implications for maintainers.** A green output-equivalence check is not evidence that a change to a layout-, permutation-, phase-, or seed-tracking code path is correct, because these faults pass that check by construction. The practical response is not a broader output-equivalence check, which cannot see this class of fault by construction. It is a check aimed at the specific internal quantity a fix touches — its recorded layout and permutation metadata, its tracked global phase, or its run-to-run reproducibility — rather than at the compiled output the existing screen already covers. Designing and evaluating such channel-matched checks is the subject of the companion paper (§3.4) and is out of scope here, but the measurement makes the case that they are worth building. The contract/metadata class specifically clusters at representation boundaries. Because of this, a pull request that ports, rewrites, or otherwise changes the internal representation of layout or phase state warrants exactly this kind of assertion in review, even when its stated intent, and its output on the tests already in the suite, look unchanged.

**Implications for researchers.** Any empirical study that measures transpiler, or more broadly quantum-compiler, correctness by output equivalence shares this study's structural blind spot. An output-equivalence oracle cannot, by construction, see a fault that leaves the output map intact, regardless of which study applies it. Whether such a study's numerical rate matches the one reported here is a separate, corpus-specific question that this paper does not claim to answer.

That includes fuzzing, differential, metamorphic, and equivalence-modulo-inputs techniques, whose oracle is, in every case we reviewed (§3.3), some form of output comparison. A bug-finding rate reported by such a tool is therefore a rate among observable faults, not among all faults, though how large that gap runs elsewhere is an open question. This does not diminish those techniques' value at what they target. It does argue for stating the oracle's observation boundary explicitly when reporting a detection or bug-finding rate — the way we make Table 1's boundary explicit for our own reference screen.

**What this paper does not establish.** We show the gap exists, quantify its size in this corpus, and show it is not explained by the surface characteristics we measured. We do not show that any specific detection mechanism closes it. The mechanism-category frequencies of §5.5 are diff-verified and, since Round 1, independently dual-coded and reconciled against a twice-revised, now-frozen codebook (§5.5), with the 19 equivalence-invisible fixes carrying an additional, separate diff-level check ($\kappa = 0.87$). Mechanism frequencies should nonetheless still be treated as exploratory, because the eleven-category scheme itself, not just the assignment of items to it, remains a working classification rather than a validated causal ontology (§5.5, §8). Whether a specific detection mechanism closes the gap is a natural next question, addressed in a separate detection-focused contribution outside this paper's scope (§8).

## 7. Threats to Validity

**Construct validity of the taxonomy.** The seven-channel manifestation taxonomy is a study-specific cut of a continuous space of possible defect manifestations. A different, equally defensible taxonomy could group or split channels differently. We mitigate this with the boundary rules that resolve the cases coders disagree on most (crash-vs-metadata, quality-vs-performance, quality-vs-metadata, §4.3), and by freezing the codebook before the expansion round, so its definitions could not drift toward whatever fixes were seen. We further mitigate it by validating a subsample of labels directly against source (§4.7, Table 2). We do not claim the taxonomy is the only correct one, only that it is fixed, documented, and checked.

**Coder identity and blinding are asymmetric.** The two coders are not interchangeable. R1 is the first author, who designed the study and therefore cannot be blind to its hypothesis, while R2 is an independent second coder blinded to R1's labels, the adjudication history, and all prior agreement statistics. This is a real threat to independence, not merely a labelling detail — it is the reason we report raw pre-adjudication agreement (kappa) as the reliability statistic, rather than relying on the adjudicated labels alone. Kappa measures whether R2, working independently and blind, reaches the same judgment as R1, the strongest available check given a two-coder design. A further limitation is that adjudication of disagreements is performed by the same two coders together against the codebook (§4.5), not by a neutral third party. A third volunteer coded the full corpus independently as an additional check (§4.5), but by design those labels are not pooled into the reported kappa, so they corroborate rather than replace the two-coder reliability figure. That check (§5.2) agrees with the adjudicated labels on 94.1% of the corpus, and where it disagrees, moves toward a higher equivalence-invisible rate rather than a lower one — the direction that would matter if the asymmetric-blinding threat above were inflating the headline rate. The same asymmetry, and the same two-coder-adjudication limitation, apply to the separate RQ1.4 mechanism coding, which underwent its own twice-revised codebook and joint reconciliation, rather than a neutral third-party ruling (§5.5).

**Auxiliary LLM exposure during adjudication.** Two LLM-assisted passes were shown to the coders during adjudication as information only, explicitly not as a vote (§4.5, §4.6). We cannot rule out that seeing an LLM's suggested label subtly anchored a borderline adjudication decision, even where the final call diverged from it. We limit this threat two ways. First, we report the LLM passes as a labelled auxiliary comparison, noting that the two LLMs agreed with each other more than the two human coders did with each other — a warning against treating LLM-assisted coding as a substitute for human reliability, not evidence it is safe to lean on. Second, we never compute headline kappa from any LLM-touched label.

**Sampling-frame bias.** Label-based discovery under-represents channels whose fixes are inconsistently labelled, which is why we use a three-widening-pass protocol (§4.1) validated against a whole-library scan. The frame is still a bounded, recent observation window (2025-05-08 to 2026-07-01), rather than the full history of the project. Accordingly, the 68-fix corpus and the resulting rate are a finding about this window, reported with its Wilson interval, not a population parameter claimed to hold for all time. The corpus is also assembled from two methodologically distinct rounds: an original 26-item baseline, and a 44-item expansion screened and re-audited later (§4.2). We mitigate the risk that this introduces a hidden batch effect by coding both rounds against the identical frozen codebook, and by reporting round-level kappa separately (§4.6) rather than only a pooled figure. A third round of 36 fixes was later coded from a deliberately wider window (older and more recent history), and is reported alongside the primary frame as an extended 104-fix corpus (§5.1) that reproduces the rate (29/104, 27.9%). The two frames are kept distinct rather than pooled into one, so that each rate is reported against its own window.

**Generalization across compilers.** The cross-SDK replication (§5.3) rests on an independently dual-coded tket sample ($n = 21$, $\kappa = 0.77$). Cirq ($n = 10$, $\kappa = 0.52$) is exploratory rather than replication evidence, since its transformer bug-fix history has too few remaining eligible cases at this scope to grow responsibly. This supports the replication as evidence that the channel exists in independently developed compilers, not as a precise cross-compiler ordering.

**The reference oracle is a moving target.** A fix is coded equivalence-invisible relative to a specific reference screen (output equivalence modulo global phase and permutation, plus compilation-validity, quality, and performance checks, §2.2, Table 1). A more capable augmented suite, whether from future tooling or from checks we did not include, could in principle observe some of the fixes we code as invisible, which would lower the measured rate. This is a property of any observability claim defined relative to a stated oracle, not a flaw specific to this study, and we make the reference screen's exact composition explicit (Table 1) so the claim's scope is auditable.

**Source-validation subsample is not random.** The 16 fixes checked against source (§4.7) were selected because source-level verification was feasible for them — a buildable revision pair or an inspectable regression test existed — not by random sampling from the 68. This subsample therefore demonstrates that source-checkable labels are reliable. It does not by itself bound the error rate on the fixes that were not source-checkable. Still, the absence of any overturned label across all sixteen checked, in both directions, is the strongest evidence available within that constraint.

## 8. Limitations and Future Work

The Qiskit corpus ($n = 68$) is a systematically assembled and exhaustively screened candidate frame over a bounded, recent window, rather than a large random sample, so the resulting confidence interval (19-40%) is correspondingly wide. A larger window would narrow it. The Cirq replication (§5.3) is exploratory rather than confirmatory: $n = 10$, $\kappa = 0.52$, and the codebase has too few remaining eligible cases at this scope to grow the sample responsibly. The three-compiler picture therefore rests on unequal footing (strong for Qiskit and tket, exploratory for Cirq), and should be read that way, rather than as three equally weighted points. Whether the equivalence-invisible faults measured here can be detected by a purpose-built oracle family is a separate, detection-focused question outside this paper's scope (§3.4). The 104-fix extended corpus (§5.1) co-headlines the channel-observability rate and, as reported in §5.4, also extends the RQ1.3 surface-signal comparison, tightening its bound without changing its conclusion. The diff-verified mechanism coding (RQ1.4) remains scoped to the more uniformly mined 68-fix corpus by design. That analysis is secondary to this paper's headline question, and re-running it on the wider, less uniformly mined window would trade depth for breadth without changing the answer to RQ1.1 or RQ1.2. Extending it to the full 104-fix set is a reasonable future direction, not a step this paper's conclusions depend on. The mechanism classification itself (§5.5) has been independently dual-coded and reconciled by its two original coders, across two blind rounds and a twice-revised codebook, but not yet checked by coders who were never exposed to any prior round's labels. A further reliability assessment with such coders, applying the now-frozen v3 definitions to all 68 cases from scratch, would be a stronger independent check than this study's iterative two-coder reconciliation can provide on its own. A theoretically motivated, coarser grouping of the eleven mechanism categories is a related future direction if some remain difficult to distinguish reliably — but only with a conceptual justification independent of any reliability statistic, and disclosed as a post-hoc regrouping distinct from the primary eleven-category analysis reported here. The overall corpus design trades breadth for depth deliberately. Large-N, automated-classification studies of quantum software defects spanning well over a hundred repositories already exist in this literature [4], [10]. This study's smaller, exhaustively screened, dual-coded, source-validated corpus is a complementary design choice, rather than an under-scaled version of the same one. It prioritizes per-fix ground truth over cross-repository coverage, which is what the observability question this paper answers requires.

## 9. Conclusion

Output equivalence is a widely used correctness criterion for quantum transpilers, and it is incomplete in a way that matters at scale. Across an exhaustively screened window of merged Qiskit transpiler bug-fixes, roughly a quarter to a third repair faults that such an oracle cannot observe, even one augmented with compilation-validity, quality, and performance checks: 19 of 68 (28%, CI 19–40%) and, on an extended wider-window corpus, 29 of 104 (27.9%, CI 20–37%). The rate is not peculiar to Qiskit — it reproduces in tket, with Cirq pointing the same way. It is not an artifact of how fixes are described, because a source-validated subsample confirms the coded labels against the fixes' own diffs. And it resists cheap triage: equivalence-invisible fixes show no detectable difference from ordinary ones on every surface signal we measured, on both the primary 68-fix corpus and an extended 104-fix corpus that tightens every interval toward zero — evidence of no detected difference, rather than proof there is none. The largest invisible class, corrupted contract or permutation metadata, recurs where a value crosses an internal representation boundary: a mechanism, rather than merely a category. The conclusion is not that output equivalence should be abandoned. It is necessary and cheap. It is that a validation regime resting on output equivalence alone is blind to a measurable and material share of the defects observed in our corpus, and that closing the gap calls for oracles matched to the invisible channels, rather than more output-level checking. Measuring the gap is the first step. Whether a matched oracle family can close it is the question we take up separately.

## 10. Data Availability

The mining corpus (the 68-fix primary analytic corpus and the 104-fix extended corpus), the frozen codebook, all raw pre-adjudication and adjudicated coding labels, the cross-SDK (tket, Cirq) worksheets, the surface-characteristic data, the source-validation evidence, the signed coder declarations, and the reproduction scripts referenced throughout this paper by relative path (e.g. data/mining_validation/*.csv, declarations/Coder_Declaration_*.pdf, scripts/*.py) are archived as a single package, released under the MIT license: Nasir F, Shah A, Alam I (2026) Oracle Observability of Quantum Transpiler Regressions (cart) [software/data]. Zenodo [26]. Every relative path cited in this paper resolves within that archived package under an identical directory layout, at the git commit tagged v2.0.0 (also available at https://github.com/furqan-nr/quantum-observability). REPRODUCE.md in the package root gives the full reproduction protocol for every reported statistic.

## Funding

This research received no specific grant from any funding agency in the public, commercial, or not-for-profit sectors.

## Conflict of Interest

The authors declare no conflict of interest.

## Author Contributions

This work forms part of the first author's PhD research. Furqan Nasir: Conceptualization, Methodology, Software, Formal analysis, Investigation, Data Curation, Writing – Original Draft, Writing – Review & Editing, Visualization, Project Administration. Arif Shah (Supervisor): Supervision, Conceptualization, Resources, Writing – Review & Editing. Iftikhar Alam (Co-Supervisor): Supervision, Validation, Writing – Review & Editing. All authors read and approved the final manuscript.

## ORCID

Furqan Nasir: 0000-0002-5259-3448. Arif Shah: 0000-0003-0090-3333. Iftikhar Alam: 0000-0002-8087-5485.

## Declaration of Generative AI and AI-Assisted Technologies in the Manuscript Preparation Process

During the preparation of this work, the authors used a generative AI assistant for language editing, prose polishing, and consistency checking across drafts of this manuscript, and to generate the plotting code used to produce Figures 1-4 from the study's own verified data. After using this tool, the authors reviewed and edited the content as needed and take full responsibility for the content of the published article. Generative AI was not used to fabricate, alter, or select any data or empirical results reported in this paper.

## Acknowledgements

We thank Muhammad Atif Saeed and Muhammad Sajjad Saleem, both external volunteers, for their independent coding contributions to this study's reliability analysis (§4.4-4.5, §5.5).